\documentclass[twocolumn,showpacs,preprintnumbers,amsmath,amssymb,prb,superscriptaddress]{revtex4}

\usepackage{graphicx}
\usepackage{dcolumn}
\usepackage{bm}

\begin{document}

\title{Reversible self-assembly of patchy particles into monodisperse icosahedral clusters}

\author{Alex W.~Wilber}
\affiliation{Physical and Theoretical Chemistry Laboratory, University of
  Oxford, South Parks Road, Oxford OX1 3QZ, United Kingdom}
\author{Jonathan P.~K.~Doye}
\thanks{Authors for correspondence}
\affiliation{Physical and Theoretical Chemistry Laboratory, University of
  Oxford, South Parks Road, Oxford OX1 3QZ, United Kingdom}
\author{Ard A.~Louis$^*$}
\affiliation{Theoretical Physics, University of Oxford, Keble Road, Oxford OX1 3NP, United Kingdom}
\author{Eva G.~Noya}
\affiliation{Departamento de Qu\'{\i}mica-F\'{\i}sica, Facultad de
        Ciencias Qu\'{\i}micas, Universidad Complutense de Madrid,
        E-28040 Madrid, Spain}
\author{Mark A. Miller}
\affiliation{Department of Chemistry, University of Cambridge, Lensfield Road, Cambridge CB2 1EW, United Kingdom}
\author{Pauline Wong}
\affiliation{Department of Radiology, University of Cambridge, 
Addenbrookes Hospital, Cambridge CB2 2QQ, United Kingdom}

\date{\today}

\begin{abstract}
We systematically study the design of simple patchy sphere models that
reversibly self-assemble into monodisperse icosahedral clusters. We find that the optimal
patch width is a compromise between structural specificity (the patches
must be narrow enough to energetically select the desired clusters) and
kinetic accessibility (they must be sufficiently wide to avoid kinetic
traps). Similarly, for good yields the temperature must be low enough for the clusters
to be thermodynamically stable, but the clusters must also have enough thermal energy
to allow incorrectly formed bonds to be broken.
Ordered clusters can form through a number of different dynamic
pathways, including direct nucleation and indirect pathways involving
large disordered intermediates. The latter pathway is related to
a reentrant liquid-to-gas transition that occurs for intermediate patch widths upon lowering the temperature.
We also find that the assembly process is robust to inaccurate patch placement up to a certain
threshold, and that it is possible to replace the five discrete patches with a
single ring patch with no significant loss in yield.

\end{abstract}


\maketitle

\section{Introduction}

The remarkable ability of biological matter to robustly self-assemble
into well defined composite objects excites the imagination,
suggesting that these processes could perhaps be emulated through the
judicious design of synthetic building blocks.\cite{Whitesides02b}
Viruses provide a particularly inspiring example. As first shown 40 years ago for the cowpea chlorotic mottle virus,\cite{Bancroft67}
and later for a wide variety of other species,\cite{Salunka89, Rombaut90, Prevelige93}
empty spherical virus shells (capsids) can be made to reversibly self-assemble
from individual protein subunits (capsomers) {\em in vitro}, simply by changing
solution conditions such as the pH. Although this process resembles micellar
self-assembly, there is an important difference: in contrast to the polydisperse
distributions that characterise micelles, the complete virus capsids are monodisperse.

Of those viruses with a roughly spherical structure, the vast majority are
found to have icosahedral symmetry. Many of the structures observed were first
predicted on symmetry grounds by Caspar and Klug.\cite{CasparKlug} More recently,
Zandi et al.\ \cite{Zandi04} and Glotzer et al.\ \cite{Glotzer07} found that the most
stable sizes of simple model systems correlated with many of the structures
found in virus capsids. 

By contrast, the mechanisms by which capsid self-assembly takes place are incompletely
understood, though there has recently been considerable theoretical 
\cite{Zlotnick05_TheoreticalAspects, Endres05, Zandi06, Hemberg06, Hicks06, Rapaport04, Hagan06, Endres02, Johnson05, Brooks06, Zhang06, Zlotnick07}
and experimental \cite{Parent05, Parent06, Parent07, Zlotnick03, Johnson04, Casini04, Teschke03} progress. 
Some common conclusions emerge from this body of work.
Firstly, one of the major potential
obstacles to successful assembly results from a class of kinetic traps, in which a large number
of partial capsids are formed but are then unable to proceed to completion because of a
lack of remaining free capsomers. This situation can be avoided if the free energy barrier for nucleation
is large, such that intermediates are formed only slowly and free capsomers remain plentiful.
\cite{Endres02, Johnson05, Brooks06, Zhang06, Zlotnick07, Parent05, Parent06, Parent07}
Secondly, it has been shown that only weak capsomer-capsomer interactions are required for capsids to be
stable, because of the large number of interactions present in a complete capsid.
\cite{Zhang06, Zlotnick07, Parent05, Parent06, Parent07, Zlotnick03} It is worth noting that weak interactions
will tend to result in a large nucleation barrier, and also allow easier breakdown of incorrectly formed intermediates. These factors
together imply that in general weak interactions will favour successful assembly, so long as they
remain strong enough for the complete capsid to be stable and for nucleation to occur in a reasonable time scale.

Understanding how self assembly with virus-like control and fidelity could be achieved with
synthetically produced sub-units is an important goal for
nanofabrication.  New experimental techniques to create
self-assembling systems are being rapidly developed. \cite{Whitesides02b}
For example, one particularly impressive recent example is the reversible formation of
tetrahedra from four appropriately designed DNA strands. \cite{Goodman05}
For colloidal and nanoparticle building blocks,
one of the most crucial requirements to allow increased control over assembly is the
synthesis of anisotropic ``patchy'' particles. There has recently been a great deal
of work in this area, with notable successes for both colloids
\cite{vanBlaaderen06,Li05,Cho05_Bidisperse,Roh05,Snyder05} and polymer-coated nanoparticles.
\cite{Jackson04,Stellacci07,Levy06c}

Both these experimental advances and the importance of understanding capsid
formation have stimulated a number of simulation studies of the self-assembly
of monodisperse objects. Rapaport \cite{Rapaport04} produced the first simulation study of capsid assembly, and
used a model consisting of trapezoidal
particles with attractive sites which assemble into a virus-like
structure. However, Rapaport's model uses several unphysical rules to assist in correct assembly,
such as irreversible bonding if subunits
come together in the correct conformation.

Subsequently, there have been a number of simulations that have
achieved reversible self-assembly. \cite{Hagan06,vanWorkum06,Zhang04,Glotzer04,Brooks06}
For example, Hagan and Chandler were able to assemble large
60-particle virus models by using as their basic ``capsomer'' units spherical
particles with directional anisotropic interactions that are chosen to be
complementary in order to help guide the particles into the right local relative orientations. \cite{Hagan06}
This feature may naturally be realised in protein-protein interactions but may be
more difficult to implement in synthetic systems. Van Workum and
Douglas also found that virus-like assemblies (not necessarily monodisperse)
formed from particles made up of three dipolar Stockmayer particles connected
together to form triangles. \cite{vanWorkum06} Zhang and Glotzer created
anisotropic model particles by rigidly connecting spheres with different
attractive potentials, and showed that these can form a rich variety of small
cluster shapes and extended structures. \cite{Zhang04,Glotzer04} Most recently 
Nguyen et al.\ performed extensive studies using a geometric model in which
appropriately shaped particles with specific interactions reversibly assembled
into icosahedral capsids. \cite{Brooks06} They produced a phase diagram for the model,
and also observed incorrectly formed ``monster'' particles very similar to those seen
with real viruses.

In this paper we describe a set of minimal models, single spheres with
anisotropic ``patchy'' interactions, that exhibit reversible self-assembly
into monodisperse cluster phases. Choosing such simple systems facilitates
the systematic exploration of parameter space to uncover the optimal design
rules for self-assembly, and helps untangle the roles of thermodynamic and
kinetic factors. This model may offer insight into how
synthetic nanoparticles and colloids could be designed to self-assemble, and while the
model is too simplistic to accurately describe the interactions between the capsid proteins of 
viruses, we hope that the principles identified will also be of relevance to biological
examples of monodisperse self-assembly.

\section{Methods}

\subsection{Model}

Our model consists of spherical particles with a number of patches whose
geometry is specified by a set of patch vectors. The repulsion is based
on the isotropic Lennard-Jones potential

\begin{equation}\label{eqn:LJ} V_{\rm LJ}(r) = 4\epsilon\left[ \left( \frac{\sigma_{LJ}}{r}
    \right)^{12} - \left( \frac{\sigma_{LJ}}{r} \right)^{6} \right], \end{equation}
but the attraction is modulated by an orientationally dependent term,
$V_{\rm ang}$.  Thus, the complete potential is
\begin{equation}
\label{eq:potential}
V_{ij}({\mathbf r_{ij}},{\mathbf \Omega_i},{\mathbf \Omega_j})=\left\{
    \begin{array}{ll}
       V_{\rm LJ}(r_{ij}) & r<\sigma_{\rm LJ} \\
       V_{\rm LJ}(r_{ij})
       V_{\rm ang}({\mathbf {\hat r}_{ij}},{\mathbf \Omega_i},{\mathbf \Omega_j})
                       & r\ge \sigma_{\rm LJ}, \end{array} \right.
\end{equation}
where ${\mathbf \Omega_i}$ is the orientation of particle $i$.
$V_{\rm ang}$ has the form:
\begin{eqnarray}
V_{\rm ang}({\mathbf {\hat r}_{ij}},{\mathbf \Omega_i},{\mathbf \Omega_j})&=&
G_{ij}({\mathbf {\hat r}_{ij}},{\mathbf \Omega_i})
G_{ji}({\mathbf {\hat r}_{ji}},{\mathbf \Omega_j})\\
\label{eqn:AngMod}
G_{ij}({\mathbf {\hat r}_{ij}},{\mathbf \Omega_i})&=&
\exp\left(-{\theta_{k_{\rm min}ij}^2\over 2\sigma^2}\right),
\end{eqnarray}
where $\sigma$  
is the standard deviation of the Gaussian,
$\theta_{kij}$ is the angle between patch vector $k$ on particle $i$ 
and the interparticle vector $\mathbf r_{ij}$,
and $k_{\rm min}$ is the patch that minimizes the magnitude of this angle.
Hence, only the patches on each particle that are closest to the interparticle
axis interact with each other.

When using this potential in our Monte Carlo
simulations, we truncate and shift the potential using a cutoff distance of 
3\,$\sigma_{LJ}$ in order to improve the computational efficiency. To maintain the 
isotropic nature of the repulsion, we also adjust the distance at which we start
to include the angular modulation so that it corresponds to where the cut-and-shifted
Lennard-Jones potential passes through zero.

The target structure we choose to try to assemble is the 12-particle hollow icosahedron
(Fig.\ \ref{fig:StructureAndGmin}(b)). This structure is sufficiently
complex to allow interesting behaviour to be observed, but forms easily enough
for large numbers of simulations of acceptable length to be carried out. It is also
analogous to the structure of ``$T=1$'' viruses.

In order to design the patchy particles so that they self-assemble into hollow
icosahedra, the natural choice for the patch positions is to place them such that
they point directly at the neighbouring particles in the target structure. The
resulting particle design, along with an assembly of twelve such particles into the target structure, is shown
in Fig.\ \ref{fig:StructureAndGmin}(b). Having chosen the patch positions, the only
parameter in the potential that remains to be optimized is then the patch width $\sigma$.

\subsection{Monte Carlo simulation}
\label{sec:MonteCarlo}

We simulate the system in the canonical ensemble using a Metropolis Monte Carlo algorithm where
the allowed move types are translations and rotations of individual particles.
The translational moves are randomly chosen from a small cube centred on the
selected particle. Rotational moves make use of a quaternion representation of the
particle's orientation, which is modified by addition of a smaller random quaternion and
then renormalized.\cite{FrenkelSmitQuaternions}

This algorithm does not directly simulate physical trajectories, and is
more often used to obtain thermodynamic averages of properties of a
system. However, while the dynamics generated are not strictly rigorous, the
use of small translational and rotational steps will mimic diffusive
dynamics.\cite{Kikuchi91,Tiana07,Berthier07} The advantage of this technique is that the derivatives of the
potential are not needed, which allows the simulations to run very efficiently.

In the analysis of the results it is useful to assign the particles to
clusters. The condition used is that two particles are considered
bonded if their interaction energy is less than $-0.4$\,$\epsilon$. Two particles 
are then considered to be in the same cluster if they are joined by a continuous chain of bonds.

As the above Monte Carlo simulations generally do not reach equilibrium, to determine some of the thermodynamic
properties we also used the umbrella sampling technique. \cite{Valleau, FrenkelSmitUmbrellaSampling} In umbrella
sampling rather than generating configurations with the Boltzmann distribution, configurations are instead sampled
according to the distribution $\exp\left[-\beta\left(\mathcal{V}+w(Q)\right)\right]$, where $w(Q)$ is a weighting function
that is a function of an order parameter $Q$. Generally, the reason for introducing the weighting function is to
artificially reduce the free energy barrier between
different (meta)stable states of a system, hence facilitating interchange
between these states and allowing the system to approach equilibrium. Canonical averages are simple to
obtain from a simulation of such a non-Boltzmann (nB) ensemble using the expression
\begin{equation} \left<B\right>_{NVT} = \left<B \exp \left[ \beta w(Q) \right] \right>_{\mathrm{nB}}, \end{equation}
where $B$ is some generic property of the system.

In our case, we wish to locate the temperature at which the system becomes most stable as a collection
of icosahedra, and so we need to overcome the free energy barrier for the formation of this structure. As the
icosahedra only very weakly interact with each other, it is sufficient, to a first approximation,
just to consider a 12-particle system.
For an order parameter we simply use the number of particles in the largest cluster present. We wish to choose
the weight function so that the system spends an approximately equal amount of time at each value of $Q$. To
achieve this we use an iterative scheme in which $w(Q)$ is updated at the end of each of a series of simulations.

\section{Results}

\subsection{Energetics}

\begin{figure}
\includegraphics[width=8.4cm]{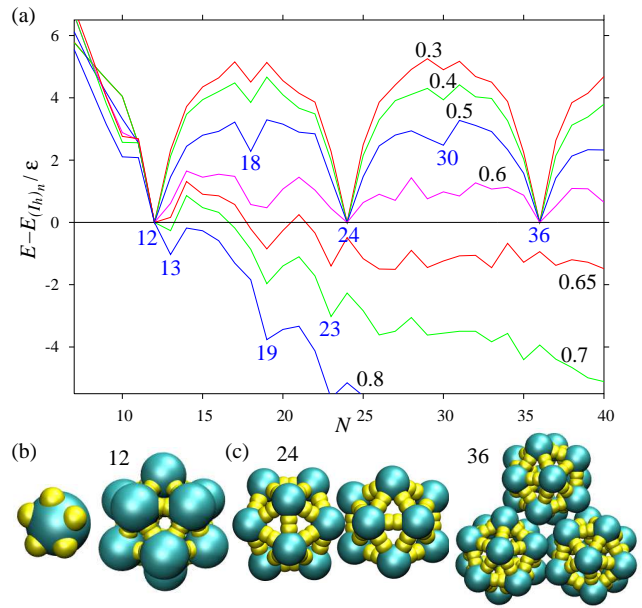}
\caption{\label{fig:StructureAndGmin}(Colour Online) (a) The size dependence of the
energy of the global minimum for different values of $\sigma$.
Each line is labelled by the value of $\sigma$ (measured in radians), with
particularly stable cluster sizes also labelled.
The energy zero is a fit to the energies of the assemblies of complete
icosahedra possible at $N$=12, 24, 36 and 48 at the appropriate value
of $\sigma$.
Hence, negative values indicate that it will never be energetically
favourable for the system to form monodisperse icosahedra. (b) A single
particle designed to form a hollow icosahedron, and a complete icosahedral cluster.
The yellow spheres represent the positions of the patch vectors.
(c) The minimum-energy structures for 24 and 36 particles for $\sigma \leq 0.6$.
The radius of the particles is reduced compared to (b) to allow the structures to be seen more clearly.
}
\end{figure}

A minimal requirement for self-assembly is that the system of
monodisperse clusters is energetically most stable, i.e.\ a
system of $n$ icosahedra must be lower in energy than the lowest-energy
single $12n$-particle cluster. We check for what range of $\sigma$ this 
requirement holds, by performing global optimisation to find the
lowest-energy configurations of our system as a function of the number of particles.
The global optimisation algorithm we used was the basin hopping algorithm, \cite{WalesD97}
which has proved to be particularly successful for clusters. \cite{WalesS99}
Fig.\ \ref{fig:StructureAndGmin} shows that for wider patches the global minima correspond to
single compact clusters with magic numbers as for the isotropic Lennard-Jones
potential ($N$=13, 19, 23, \dots). \cite{Northby87} However,
for sufficiently narrow patches ($\sigma \leq 0.6$) the most
stable sizes occur at $N$=12, 24, 36, \dots, and do indeed correspond to
packings of 1, 2, 3, \dots discrete icosahedra (Fig.\ \ref{fig:StructureAndGmin}(c)), as desired.
(Note that as the icosahedra weakly interact, the discrete icosahedra
are touching in the global minimum configuration).
The energy landscapes for such magic number clusters are likely to have
a single-funnel topography. \cite{Wales05, Doye99f}

\subsection{Self assembly}

\begin{figure}
\includegraphics[width=8.4cm]{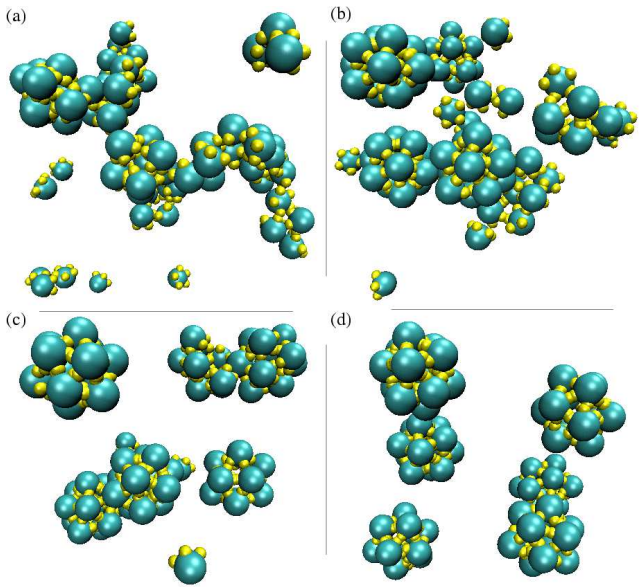}
\caption{\label{fig:snapshots}(Colour Online) Snapshots of a system
of 72 particles assembling into six complete icosahedra
at $\sigma=0.45$ and $T=0.14\,\epsilon k_B^{-1}$ after
(a) 3000, (b) 8000, (c) 60\,000 and (d) 250\,000 MC cycles.
}
\end{figure}

To further investigate the self-assembly behaviour of our system,
we performed Monte Carlo (MC) simulations in the canonical ensemble.
Fig.\ \ref{fig:snapshots} illustrates a typical trajectory, and
shows that, given the right conditions,
the patchy particles are able to reversibly assemble into a monodisperse set
of hollow icosahedral clusters.
At the beginning of the simulation, disordered clusters form rapidly 
(Fig.\ \ref{fig:snapshots}(a)). These clusters then soon begin
to order. For example, five-coordinate vertices become visible, and many
of the particles already occupy their correct positions in partially formed
capsids (Fig.\ \ref{fig:snapshots}(b)). The slowest process is then the diffusion
of the few remaining monomers to the empty sites in the clusters (Fig.\ \ref{fig:snapshots}(c) and (d)).

\begin{figure}
\includegraphics[width=8.4cm]{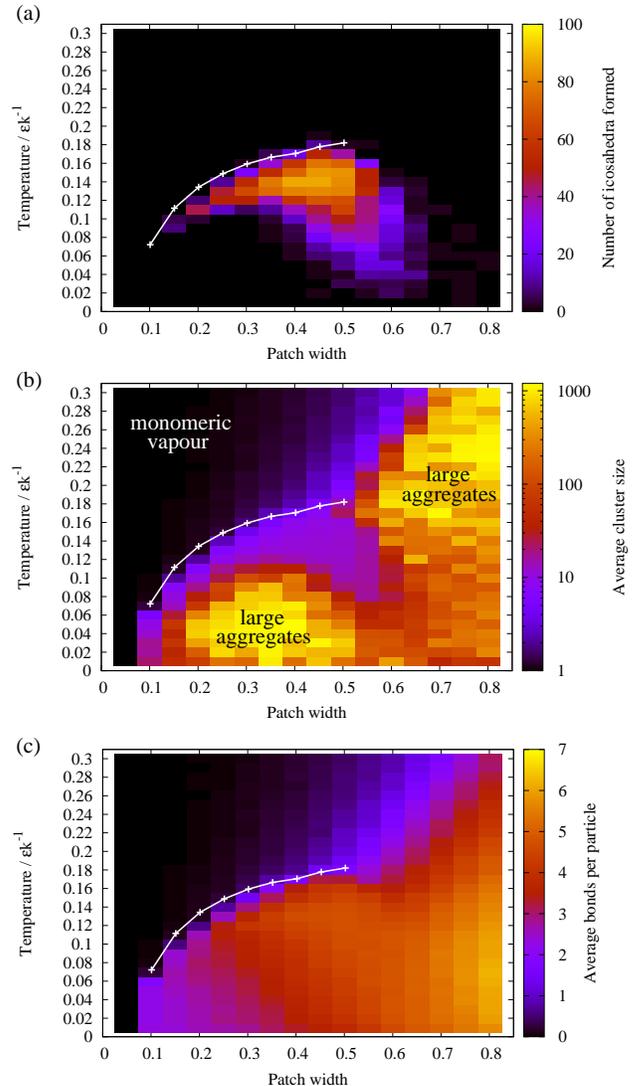}
\caption{\label{fig:MainHeatMap}(Colour Online) (a) The number of
icosahedra formed, (b) the mean cluster size (averaged over particles)
and (c) the average number of bonds per particle (counting each bond twice)
after $80\,000$ MC cycles as a function of the patch width $\sigma$ (measured
in radians) and the temperature for 1200 particles at a number density of
$0.15\,\sigma_{\rm LJ}^{-3}$. The white lines show the
thermodynamic transition temperature $T_{\mathrm{clust}}$ for the transition from
icosahedra to a gas of monomers (see Section \ref{sec:Thermo}).} 
\end{figure}

The simplicity of our model allows us to map out in detail the assembly
behaviour, thus enabling the optimal patch width and temperature to be
identified. Fig.\ \ref{fig:MainHeatMap} summarizes results from
480 simulations at different combinations of $\sigma$ and $T$, where
each simulation was started from a disordered configuration
generated at high temperature.
Although this plot is for a particular density and after a certain number
of MC moves, the generic features are not sensitive to these parameters
(for the dependence on density see Section\ \ref{sec:Density}).

The key feature of Fig.\ \ref{fig:MainHeatMap} is
that there is a limited range in the space of patch
width and temperature where self-assembly occurs efficiently.
The maximum yield of 88\% occurs at $\sigma=0.45$ and
$T=0.14\,\epsilon k_B^{-1}$. The general patterns shown in Fig.\ \ref{fig:MainHeatMap} are relatively insensitive to simulation time,
although in general yields increase with time 
For example, we have observed yields as high as 98\% after $800\,000$ 
MC cycles.

To understand the assembly kinetics summarized in Fig.\ \ref{fig:MainHeatMap},
we first examine the effect of temperature at the optimal
$\sigma$. There is a clear maximum in the number of icosahedra formed,
with the number dropping to zero at high and low temperature. The
upper limit is thermodynamic because at high temperatures the
stable phase of the system is a gas of monomers.
The lower limit
is kinetic and arises because at low temperatures the system
does not have enough thermal energy to escape from incorrect
arrangements of the particles, and so large low-density kinetic
aggregates grow. These two limits are clearly visible from the mean
cluster size plotted in Fig.\ \ref{fig:MainHeatMap}(b).

\begin{figure}
\includegraphics[width=8.4cm]{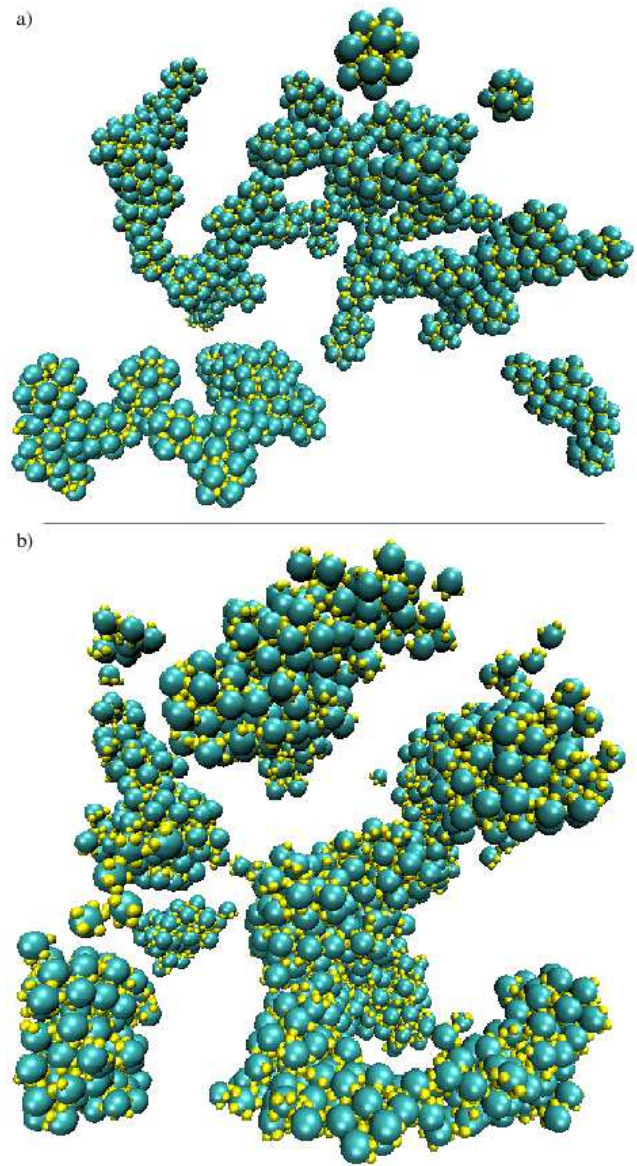}
\caption{\label{fig:Aggregates}(Colour Online) Typical aggregate structures
  formed at non-optimal temperatures and patch widths. (a) String-like
  kinetic aggregates formed at $T=0.04\epsilon k^{-1}$ and $\sigma=0.6$.
  (b) Thermodynamic liquid-like
  aggregates formed at $T=0.18\epsilon k^{-1}$ and $\sigma=0.65$.}
\end{figure}

A snapshot of
a kinetic aggregate is shown in Fig.\ \ref{fig:Aggregates}(a). This
structure is typical of those formed at higher $\sigma$ and low temperature. The
relatively wide patches allow significant rearrangement of the clusters, and
it can be seen that there is a large degree of local order, with most particles
forming strong interactions through most or all of their five patches. The diameters
of the ``strings'' in the network are roughly equal to the diameter of an ideal icosahedron.
However, the low temperature prevents complete rearrangement to form icosahedra.
For lower $\sigma$ and low $T$ the frustration of the system is much more severe, and most particles
only manage to form two or three strong interactions (Fig.\ \ref{fig:MainHeatMap}(c)). The resulting
network does not coalesce into dense clusters like
those in  Fig.\ \ref{fig:Aggregates}(a), and the ``threads'' of the network are only one
 particle in diameter. At very low $\sigma$
very few interactions are formed at all, and only small clusters are found.

Having a temperature window where ordering can occur such as that observed here is common.
For example, in the protein folding community it has been argued that
increasing the ratio $T_f/T_g$ ($T_f$ is the `folding' temperature below
which the native state becomes most stable and $T_g$ is the `glass
transition' temperature below which the dynamics becomes too slow for
folding to occur) enhances the ability of a protein to fold.
\cite{Bryngelson95} Similarly, here we find that the optimal value
of $\sigma$ occurs roughly where there is the largest difference between
$T_{\rm clust}$, the temperature below which the clusters become thermodynamically
most stable, and $T_g$, below which ordering is kinetically hindered.

The effect of reducing $\sigma$ from its optimal value is to narrow
the temperature window over which icosahedra form. Firstly, $T_{\rm
clust}$ decreases as the patches become narrower, because the
vibrational entropy of the icosahedral clusters decreases.  Secondly,
the potential becomes ``stickier''
as the patches narrow, and so the `glass transition' temperature $T_g$ below which the
particles becomes trapped in large non-equilibrium aggregates
initially increases.
Consequently, for $\sigma\le 0.1$ icosahedra are virtually never seen.

Fig.~\ref{fig:StructureAndGmin} shows that beyond a certain $\sigma$, the
icosahedra are no longer the most stable particle configuration.
This is mirrored in Fig.~\ref{fig:MainHeatMap}(b)
by the large disordered clusters that form
for larger $\sigma$. At these values of $\sigma$ and density the system phase separates
into one or more liquid droplets coexisting with a vapour. A typical configuration from
this region of the phase diagram is shown in Fig.\ \ref{fig:Aggregates}(b). Note that
the liquid droplets are not particularly spherical. The system has a low surface tension
because most patches for the particles on the surface of the droplets are pointing inwards.
As a result large shape fluctuations in the droplets are observed.

The interaction
between the clustering transition and the formation of a liquid phase gives rise to a particularly
intriguing form for the phase diagram of this system.  For example, as the temperature is lowered for $\sigma
\approx 0.55$, the system first condenses from a monomeric gas to a
bulk liquid-like phase. However, at lower temperature it forms a gas phase again,
but now made up of weakly interacting icosahedral clusters. This
reentrant lower liquid-to-vapour transition is driven by the
lower energy of the clusters, which overcomes the higher entropy of the liquid.
Preliminary parallel
tempering simulations have  confirmed this unusual scenario. The probable
implication is that there is a closed-loop liquid-vapour coexistence line in
the phase diagram for this system. This will be explored further in future work.

\subsection{Thermodynamics}
\label{sec:Thermo}

\begin{figure}
\includegraphics[width=8.4cm]{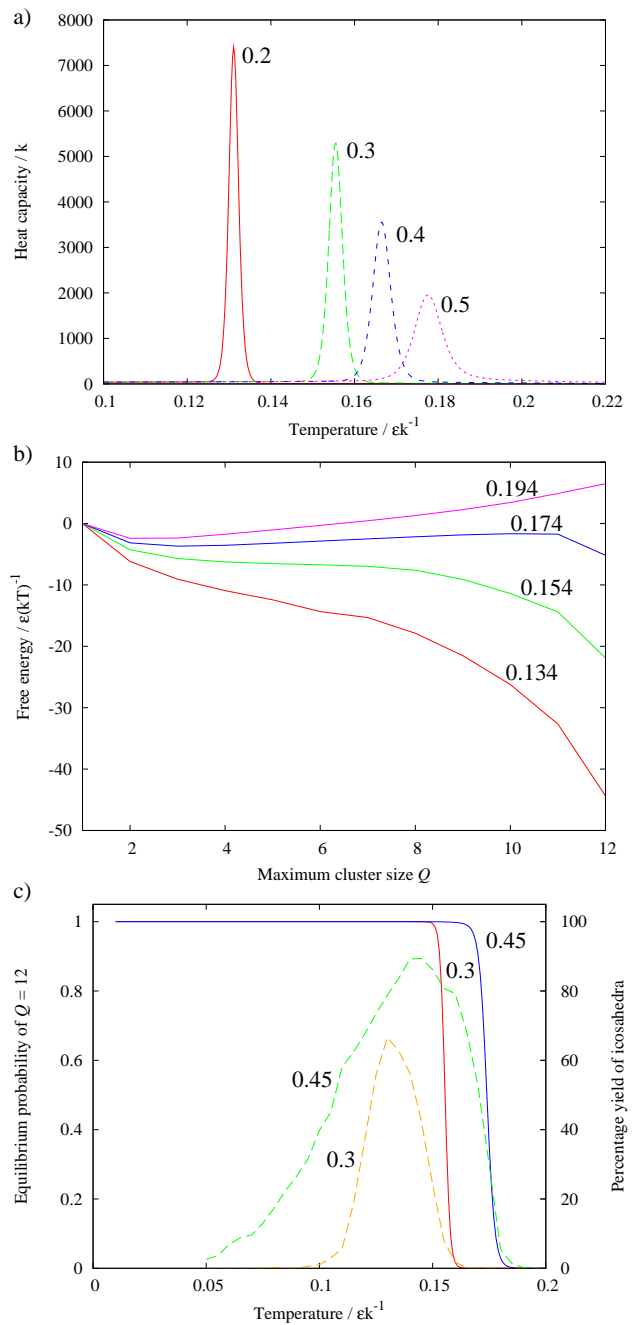}
\caption{\label{fig:ThermoPlots}(Colour Online)
Some thermodynamic properties of a 12-particle system. (a) The heat capacity as a function of temperature for different values of
$\sigma$. Each line is labelled by the value of $\sigma$ (given in radians).
(b) The free energy associated with 
the order parameter $Q$ plotted at four different temperatures, all at a patch width 
of $\sigma = 0.45$. Each line is labelled by the temperature, with $T = 0.174$
being the value of $T_{\mathrm{clust}}$ for this patch width.
(c) Equilibrium probability of $Q=12$ (solid lines). For comparison the yield of icosahedra obtained after 80\,000 MC
cycles in self assembly simulations (dotted lines) is included. Each line is labelled with the value of $\sigma$ (given in radians).}
\end{figure}

In order to obtain estimates for $T_{\mathrm{clust}}$, the temperature below which the system
is most stable as a collection of icosahedra, we made use of the umbrella sampling scheme described in Section
\ref{sec:MonteCarlo}. The resulting heat capacity plots obtained for a
selection of patch widths are shown in Fig.\ \ref{fig:ThermoPlots}(a).

The peaks in the heat capacity signal the transition from a gas of monomers to a gas of icosahedra. As an aside,
we should note that this transition is not a phase transition, because it is fundamentally a transition associated
with a finite cluster, e.g.\ the heat capacity peaks will retain the same finite width, no matter how large the system is.

The positions of the peaks provide estimates for $T_{\mathrm{clust}}$, which are shown superimposed on 
Fig.\ \ref{fig:MainHeatMap}(a). It is interesting to note that relatively little supercooling is required for the
icosahedra to begin to assemble. This is confirmed in Fig.\ \ref{fig:ThermoPlots}(c) where we compare the
equilibrium probability of a particle being in an icosahedron to that obtained in our MC
simulations of the self assembly. In particular, when the icosahedra first become stable as the temperature is decreased and
their equilibrium probability is still low, the yield in our self assembly simulations mirrors (to within
statistical error) the equilibrium results.

The free energy profiles, $A(Q)$, can be obtained straightforwardly from the umbrella sampling simulations, through
the relation:
\begin{equation} A(Q) = A - kT \ln p(Q), \end{equation}
and a series of profiles are plotted for the optimum patch width in Fig.\ \ref{fig:ThermoPlots}(b). The change in the
relative stability of the icosahedra compared to the vapour as the transition is crossed is apparent, and for sufficiently low temperature the profile becomes barrierless. At $T_{\mathrm{clust}}$,
the free energy barrier is of the order of 2\,$kT$. This relatively low value suggests that overcoming
the nucleation barrier is not rate-limiting, and helps to explain why the yield
of icosahedra initially follows the equilibrium line as the temperature decreases from above
$T_{clust}$ (Fig.\ \ref{fig:ThermoPlots}(c)). 
However, as the equilibrium population of icosahedra increases, the yield from
our self-assembly simulations begins to deviate from the equilibrium line, initially because the time scale for diffusion of free particles to 
the correct sites on incomplete clusters begins to limit the yield.

\subsection{Mechanisms of assembly}

\begin{figure}
\includegraphics[width=8.4cm]{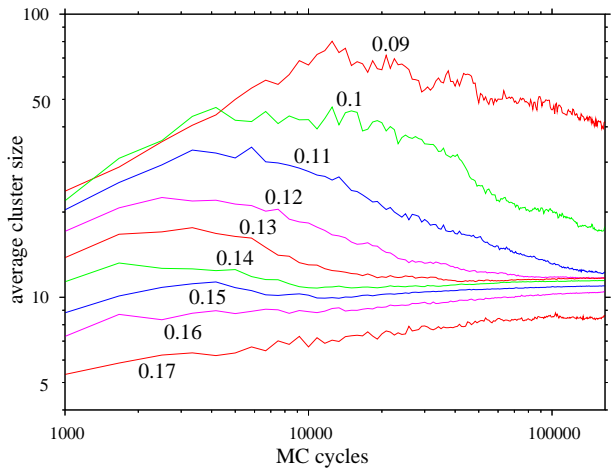}
\caption{\label{fig:dynamics}(Colour Online)
The mean cluster size as a function
of number of MC cycles at $\sigma=0.45$ and at different temperatures (as
labelled) for a system of 1200 particles at a number density of
$0.15\,\sigma_{\rm LJ}^{-3}$. Each line is an average over 10
simulations.}
\end{figure}

Further information about the mechanisms of self-assembly can be
gleaned from Fig.~\ref{fig:dynamics} which shows how the average
cluster size evolves with MC time for different temperatures at the
optimal patch width.  One striking feature is that for
$T\lesssim 0.14\,\epsilon k^{-1}$ the average cluster size goes through a
maximum before decreasing towards twelve. The pathway to cluster
formation is through larger disordered clusters, which then ``bud off''
icosahedra, rather than through the growth of icosahedra from smaller
units.  When an icosahedron with all the patches correctly oriented forms
within a larger cluster, it will only interact very weakly with the rest
of the cluster, and so will be able to escape relatively easily.
The height of the maximum in the average cluster size increases with decreasing
temperature because the rate of cluster growth increases compared to the rate
at which icosahedra are annealed out.

These results indicate therefore that the self-assembly of icosahedra
is facilitated by the formation of a metastable intermediate
phase, and the liquid phase, which we previously noted is stable at
larger $\sigma$, plays an important role, even when it has disappeared
from the equilibrium phase diagram. This system therefore provides
an example of Ostwald's step rule \cite{OstwaldStepRule} where nucleation
initially leads to a metastable phase. For example, protein
crystallization can be enhanced by the formation of a metastable
protein-rich phase from which the crystal can more easily nucleate.
\cite{tenWolde97,Vivares05}

\begin{figure}
\includegraphics[width=8.4cm]{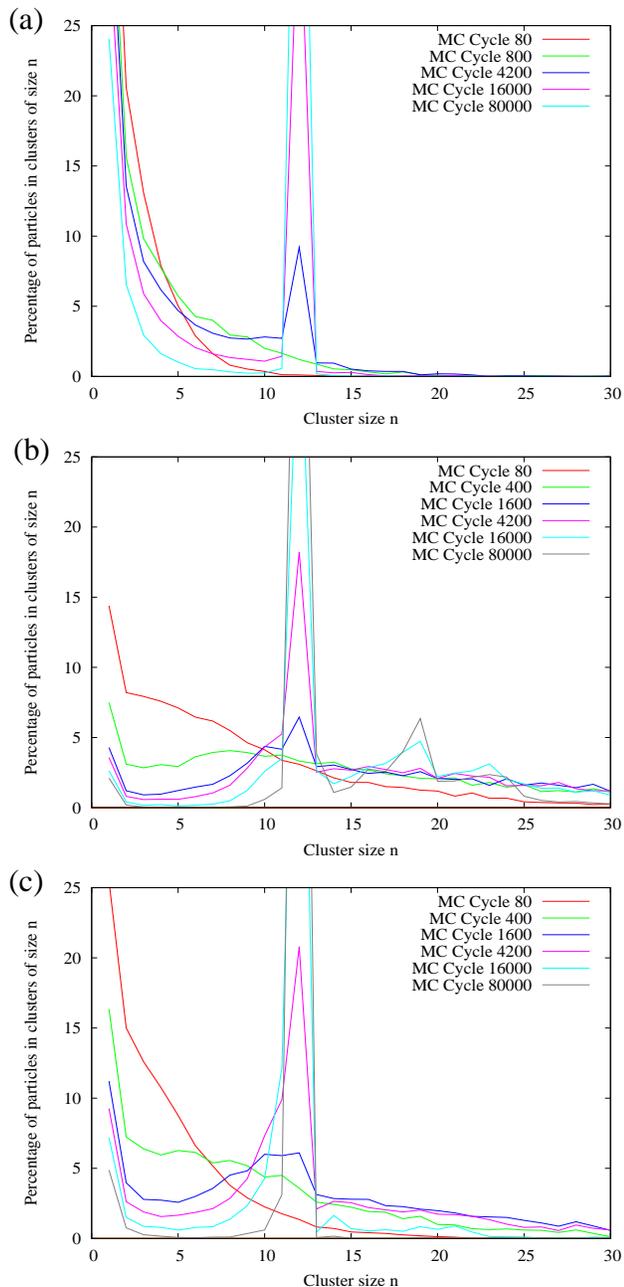}
\caption{\label{fig:ArdPlots}(Colour Online)
The fraction of particles in clusters of a certain size as a function
of cluster size at different simulation times for parameters (a) close to the transition to a monomer gas,
$T = 0.16$, $\sigma = 0.4$ (b) close to the transition to a liquid,
$T = 0.14$, $\sigma = 0.55$ and (c) giving optimal conditions for assembly, 
$T = 0.14$, $\sigma = 0.45$. In each case the data is an average
over 100 simulations, where each simulation is of 1200 particles at a number density of $0.15\,\sigma_{\rm LJ}^{-3}$.}
\end{figure}

Fig.\ \ref{fig:ArdPlots} illustrates some of the changes in assembly mechanism as a function of temperature and patch width.
In general, conditions of high temperature and low $\sigma$ favour a mechanism in
which capsids are built up by a stepwise addition of monomers and small
clusters. In Fig.\ \ref{fig:ArdPlots}(a) the cluster size distribution is dominated by monomers 
and small intermediates after 80 MC cycles. Over the course
of the simulation the population of particles shifts steadily into icosahedra, with almost no clusters larger than 12 particles being formed. 
Conversely,  conditions of low temperature and high $\sigma$ 
(close to the range of parameters where there is a stable liquid phase \cite{Tavassoli02}) favour the formation of large
clusters, which then anneal to icosahedra. Fig.\ \ref{fig:ArdPlots}(b) shows the rapid formation of very large clusters at the start of the simulation.
The average cluster size then shrinks back down by the ``budding'' mechanism mentioned above until the majority of
particles are in icosahedra. It is interesting to note that at long times the large clusters
that are left preferentially adopt sizes that are magic numbers for the isotropic Lennard-Jones
potential, e.g. the peak at $N=19$ in Fig.\ \ref{fig:ArdPlots}(b) that corresponds to two
interpenetrating centred icosahedra. \cite{Northby87} At the optimal conditions
for assembly (Fig.\ \ref{fig:ArdPlots}(c)), a combination of the two mechanisms seems to operate.

One effect of the
availability of this budding mechanism is to increase the range of temperature
over which icosahedra assemble. For example, it is noticeable from Fig.\ \ref{fig:ThermoPlots}(c)
that as $\sigma$ decreases and the system moves further from the region of
parameter space where the liquid phase is stable, the temperature range for
which there are significant yields decreases significantly.

The budding mechanism of self-assembly is very different from that so far
observed in experimental \cite{Casini04,Parent05,Zlotnick03b}
and theoretical (both kinetic models \cite{Zlotnick03b,Endres02} and direct
simulations \cite{Hagan06}) work on viruses, where direct nucleation of the
virus capsid by stepwise addition of individual capsomers is the norm. We suspect that the greater specificity of the
interactions in real viruses both makes the formation of large aggregates less
likely under conditions where the complete capsids are most stable, and
prevents the easy rearrangement of subunit
particles in any large aggregates that form.


\subsection{Dependence on density}
\label{sec:Density}
\begin{figure}
\includegraphics[width=8.4cm]{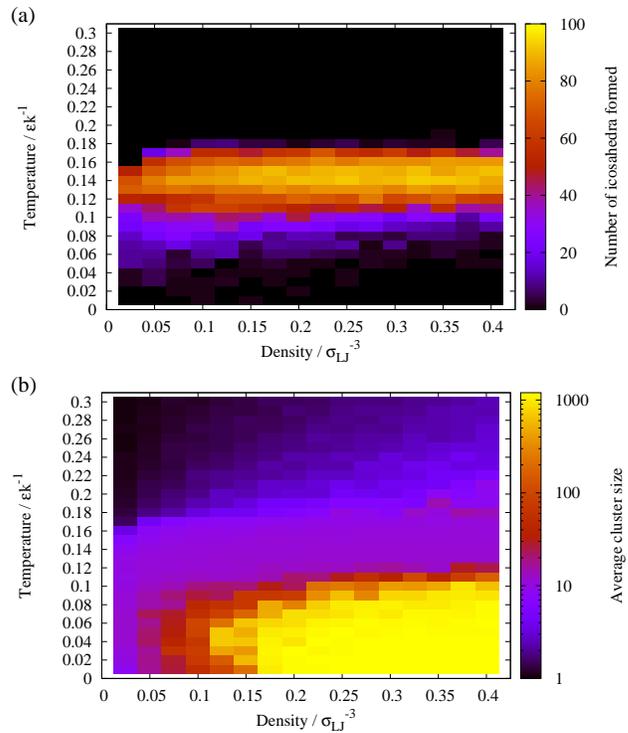}
\caption{\label{fig:DensityMaps}(Colour Online) (a) The number of icosahedra formed
  and (b) the mean cluster size (averaged over particles) after $80\,000$ MC cycles as a function of density
  and temperature for 1200 particles at a fixed patch width of $\sigma = 0.45$.}
\end{figure}

All of the results described thus far are for simulations at the same density,
$0.15\,\sigma_{\rm LJ}^{-3}$, which raises the question of how the assembly behaviour depends on particle
density. Fig.\ \ref{fig:DensityMaps}(a) shows the yield of icosahedra at a range of particle
densities, for the value of $\sigma$ that we found to be optimal for 
a density of $0.15\,\sigma_{\rm LJ}^{-3}$. The dependence of the yield on density
is weak, with markedly decreased yields only observed at very
low densities, where the times required for the particles to diffuse
and reach each other are significantly longer. At very low densities, assembly also becomes comparatively
more successful at low temperatures. This small effect arises because the formation of large
aggregates, which generally competes with the assembly of icosahedra at lower temperature, is disfavoured
at low density (Fig.\ \ref{fig:DensityMaps}(b)), and so the successful growth of icosahedra by monomer addition is somewhat enhanced.

\subsection{Effect of inaccurate patch placement}

If self-assembling structures are to be constructed from
nanoparticles and colloids, experimental methods will have to be developed to pattern the
surfaces of the particles appropriately with ``sticky'' and ``non-sticky''
regions. Since a certain degree of variability is intrinsic to synthetic colloids, 
it is useful to consider how robust the self-assembly process is with
respect to imperfect placement of the patches. These factors may also be important in the
consideration of the evolution of proteins that self-assemble into some quaternary structure, as the robustness
of self-assembly will have a strong effect on how easily such proteins can evolve.

In order to examine the effect of imperfect patch placement on assembly,
simulations were performed where a random noise term was added to the patch vectors
for each particle, with the noise for each patch on each particle being
generated separately.
The modified patch vectors $\mathbf{\hat{p}}$ are given by
\begin{equation} \mathbf{\hat{p}} = \frac{\mathbf{\hat{p_0}} + \mu \mathbf{\hat{p_1}}}{\left|\mathbf{\hat{p_0}} + \mu \mathbf{\hat{p_1}}\right|} \end{equation}
where $\mathbf{\hat{p_0}}$ is the original (noiseless) patch vector, $\mathbf{\hat{p_1}}$ is a random unit
vector, and $\mu$ is a parameter that determines the magnitude of the noise in the patch
placement. The average angle of deviation of $\mathbf{\hat{p}}$ from
$\mathbf{\hat{p_0}}$ is approximately linear in $\mu$, with a proportionality constant of 0.78 radians.

\begin{figure}
\includegraphics[width=8.4cm]{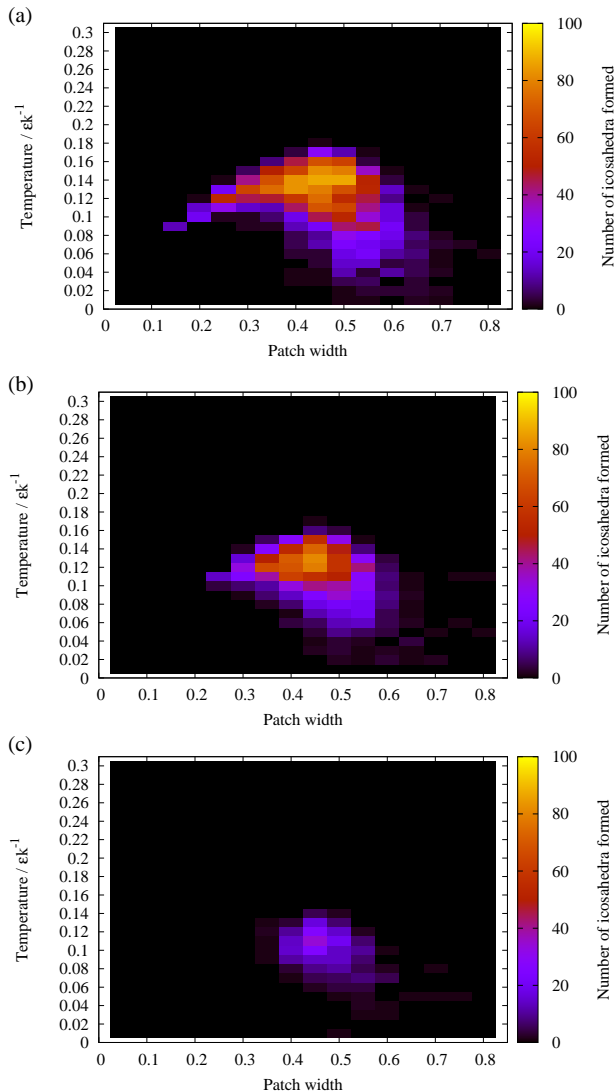}
\caption{\label{fig:NoiseMaps}(Colour Online) The effect of noisy patch placement.
  The heat maps show the numbers of icosahedra formed
  after $80\,000$ MC cycles as a function of the patch width $\sigma$
  (measured in radians) and the temperature for 1200 particles at a number
  density of $0.15\,\sigma_{\rm LJ}^{-3}$ for different noise levels in the patch placement: 
  (a) $\mu = 0.1$, (b) $\mu = 0.2$ and (c) $\mu = 0.3$.}
\end{figure}

Analogous plots to Fig.\ \ref{fig:MainHeatMap}(a) are depicted in Fig.\ \ref{fig:NoiseMaps} for increasing
values of the parameter $\mu$.
A low level of noise ($\mu=0.1$) seems to have little effect on the formation of icosahedra.
Yields away from the optimum parameters are generally slightly
depressed, and the temperature above which icosahedra are not observed is
slightly reduced. With increasing noise levels however, the region in the 
$(\sigma, T)$ plane where assembly is possible shrinks rapidly, and by 
$\mu = 0.3$ it has nearly disappeared.
The noise has the general
effect of reducing the binding energy of an icosahedral structure. Thus, the
transition temperature to a monomer gas is reduced, as is the patch width
above which larger aggregates become competitive. At low $\sigma$, i.e.\ narrow patches, the reduction
in the binding energy is more pronounced, and icosahedra again become unstable.

\begin{figure}
\includegraphics[width=8.4cm]{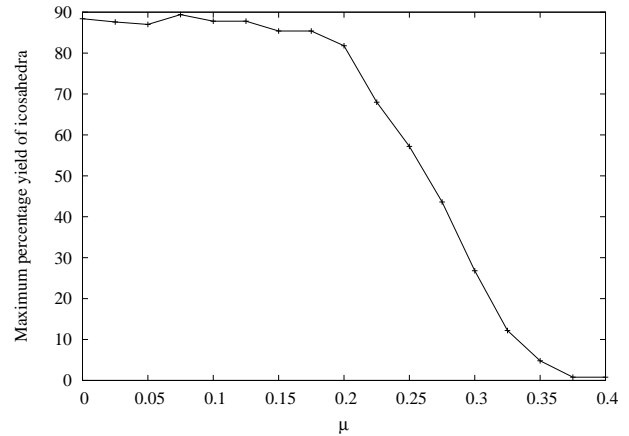}
\caption{\label{fig:NoiseYields}The maximum yield of icosahedra as a
  function of the patch placement noise parameter $\mu$. In each case the yield is
  the average of five simulations at the values of $T$ and $\sigma$ which give
  the maximum yield for that value of $\mu$. Each simulation contains 1200
  particles, at a number density of $0.15\,\sigma_{\rm LJ}^{-3}$.}
\end{figure}

The maximum yield of icosahedra as a function of $\mu$ is shown in Fig.\
\ref{fig:NoiseYields}. Interestingly, the maximum yield is virtually
independent of noise level up to $\mu = 0.2$, after which it falls off
rapidly. Beyond $\mu = 0.35$ there is virtually no successful assembly. 
$\mu = 0.2$ corresponds to a mean deviation of $0.16$ radians,
or around $9^\circ$, and may represent a target for the accuracy of patch
placement in experiments, although one would expect the threshold to be somewhat dependent on
the target structure.

\subsection{Use of a single ring patch}

The experimental production of five distinct patches that are accurately placed on a nanoparticle or colloidal surface
is likely to be very challenging. One potentially easier alternative may be the creation of
a continuous attractive ring on the particle's surface. \cite{Zhang04} To mimic the effect
of the discrete patches the ring should
pass through the positions at which the individual patches would otherwise have been
placed.

\begin{figure}
\includegraphics[width=8.4cm]{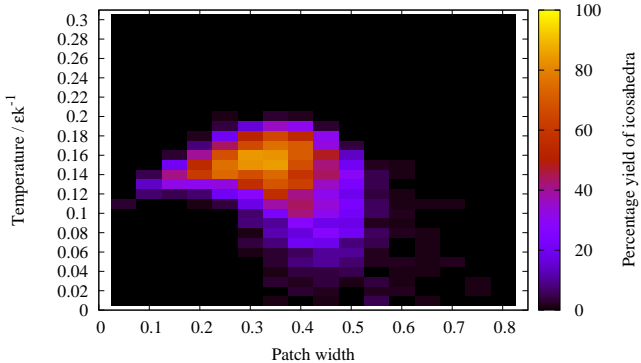}
\caption{\label{fig:RingPatches}(Colour Online) Icosahedron yields using ring patches. The 
  heat map shows the number of icosahedra
  formed after $80\,000$ MC cycles as a function of the patch width $\sigma$
  (measured in radians) and the temperature for 1200 particles at a number
  density of $0.15\,\sigma_{\rm LJ}^{-3}$, using a single ring patch on each particle
  with $\nu = 1.017$ radians.}
\end{figure}

To describe such a case the angular modulation of the interaction potential in the model (Eq.\ \ref{eqn:AngMod}) is replaced by
\begin{equation} G_{ij}\left(\hat{\mathbf{r}}_{ij},\mathbf{\Omega}_i\right) =
  \exp\left(-\frac{\left(\theta_{k_{ij}} - \nu\right)^2}{2\sigma^2}\right), \end{equation}
where the orientation of the ring patch is described by a patch vector that
passes through the centre of the ring, $\nu$ is the cone angle
from the patch vector at which the attraction is greatest, and $\sigma$ is 
a measure of the angular width of the ring.
The yields of icosahedra for particles with a single ring patch at the appropriate value of $\nu$,
namely $\nu = 1.017$ radians, are shown in Fig.\ \ref{fig:RingPatches}.

A number of comparisons can be made to the simulations where five separate
patches were used. Firstly, the maximum yield observed when using a ring patch
is the same to within statistical error (85\% as opposed to 88\% obtained with five discrete patches). It seems that the loss of
specificity in changing to a ring patch is not
too important, since the value of $\nu$ is chosen such that the most favourable
configuration is still one where each of the particles has five
neighbours. Secondly, the icosahedra are also stable to a slightly higher temperature when using
ring patches, because each particle in an icosahedron is constrained in only two orientational degrees of
freedom rather than three, reducing the entropy difference between the icosahedra and the monomeric vapour.
Thirdly, the optimal region is shifted to lower
$\sigma$, because for a given value of
$\sigma$ more of the surface of the particles is attractive than for
particles with discrete patches at the same value of $\sigma$. Consequently liquid-like
aggregates become competitive at lower values of $\sigma$. Finally, the assembly process
is somewhat less vulnerable to kinetic traps, because the continuous ring patch facilitates rearrangement
of the particles.

Our results suggest that ring patches represent an attractive alternative method of patterning
particle surfaces to form specific target structures, with prospects for
easier synthesis and comparable yields to those obtained using discrete patches.

\subsection{Effect of a short-range potential}

\begin{figure}
\includegraphics[width=8.4cm]{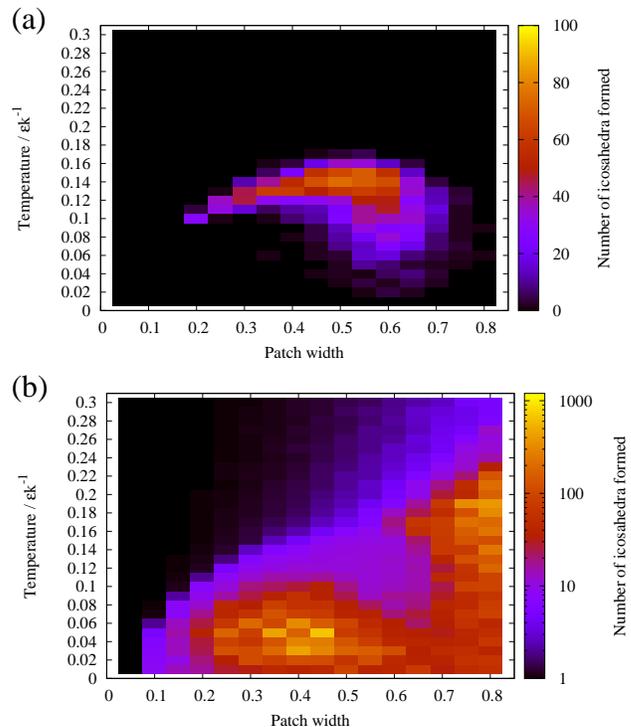}
\caption{\label{fig:SRTable20-10}(Colour Online) (a) The number of icosahedra formed
  and (b) the mean cluster size (averaged over particles) after $80\,000$ MC cycles using a
  shorter-range 20-10 potential as a function of the patch width
  $\sigma$ and temperature for 1200 particles at a number density of $0.15\,\sigma_{\rm LJ}^{-3}$.}
\end{figure}

All of the previous simulations have used an orientationally modulated version
of the Lennard-Jones potential given in Eq.\ \ref{eqn:LJ}. However, the interaction 
potentials for real colloids and proteins tend to be much shorter in range. 
\cite{Rosenbaum96} Using a longer range potential has computational advantages,
but in order to investigate the effect of the range of the potential, a set of
simulations was performed using a generalised Lennard-Jones potential with exponents of 20 and 10:
\begin{equation} V_{\rm LJ}^{20\mbox{-}10}(r) = 4\epsilon\left[ \left( \frac{\sigma_{LJ}}{r}
    \right)^{20} - \left( \frac{\sigma_{LJ}}{r} \right)^{10} \right]. \end{equation}
The results of these simulations are shown in Fig.\ \ref{fig:SRTable20-10}.

The overall trends are similar to those for the longer range potential.
However, yields are generally reduced, because the
shorter range potential makes it less likely for particles to enter each
others' attractive potential wells and form bonds, and because the barriers
to rearrangements increase. \cite{Miller99} As a result the entire
self assembly process proceeds more slowly. 
At longer simulation times the yields recover, with a maximum yield of 
icosahedron of 97\% after $1.2\times 10^6$ MC cycles.
Another general effect is that because of the slower
bond formation, any aggregates that form will tend to be reduced in size (Fig.\ \ref{fig:SRTable20-10}(b)).

The shorter range of the 20-10 potential energetically destabilises the liquid
phase, \cite{Hagen94, Doye96a} thus increasing the maximum value of $\sigma$ at
which the system is stable as a collection of icosahedra. Hence, icosahedra are
able to assemble up to higher $\sigma$. Furthermore, since the budding mechanism
of assembly requires the formation of a liquid droplet, assembly by nucleation
is also comparatively more dominant than in the case of the 12-6 potential.

Decreasing the range of the potential also has some other small effects,
seen in Fig.\ \ref{fig:SRTable20-10}(a). The vibrational entropy of the icosahedra is
reduced, lowering $T_{\mathrm{clust}}$ slightly. There is also a small increase in the value of $\sigma$
below which the kinetic suppression of icosahedron formation occurs, since at low $\sigma$ the short range further
increases the difficulty of assembly.

\section{Conclusions}

By making use of a minimal model, we have been able to simulate the behaviour of
patchy particles designed to assemble into monodisperse icosahedral clusters.
We found that, given the right conditions, the assembly process was
rapid, efficient and robust. By performing arrays of simulations at different
temperatures and using different patch widths, we were able to map out the regions
of parameter space in which assembly is successful, and to identify the major
competing behaviours arising in the regions of parameter space which did not lead to successful assembly.

From these results we can learn something about the general principles
required when designing objects to self-assemble.  The optimal patch
width represents a compromise between the energetic and kinetic
requirements for self-assembly.  As the patch width is decreased, the energy
gap between the target structure and other possible competing
structures increases.  By contrast, kinetic accessibility improves as
$\sigma$ increases, because the system is able to escape
from incorrect configurations more easily.  The lesson is that the interactions
need to be specific enough to sufficiently favour the target
structure, but that overspecifying them can inhibit the dynamics of
assembly.

We also found that the dynamic pathways to self-assembly can be
complex and non-intuitive.  We observed, for example, that, besides the
better known nucleation pathways, clusters could also form through
disordered intermediates. These mechanisms are
less likely to be important in biological systems, but may be relevant for
synthetic self-assembling systems. In our simulations the optimal yields
were found under conditions where both mechanisms made contributions to the assembly.

Assembly via nucleation pathways appears to be much more successful close to the maximum
temperature at which icosahedra are stable. Kinetic traps are avoided, and alternative processes
such as the formation of liquid droplets are inhibited. Since high temperatures are equivalent to weak
bonding interactions, this finding is consistent with the experimental observation that
protein-protein interactions in virus capsids are generally weak. \cite{Zlotnick03}

The assembly process is not strongly affected by the particle density. Changing the range of the
potential also has little effect on the yields at long times (although assembly mechanisms involving
liquid droplets are inhibited by a shorter range potential). The relative independence of the behaviour of
the system with respect to these factors indicates that our findings are quite general, and not just specific
to the parameter range we focussed on.

The self-assembly process is encouragingly robust with respect to
the design and placement of patches. Simulations in which the patches were
positioned with random inaccuracies still produced almost unchanged yields
of correctly formed icosahedra up to a threshold level of average error.
While most of our simulations used particles with five discrete patches, we
found that particles with a single ring-shaped patch, which may be easier to produce
experimentally, assembled with a similar
level of efficiency. Both of
these findings may be seen as very encouraging for experimentalists seeking to
synthesise self-assembling nanoparticles and colloids.

Our study focussed on a simple model in order to be able to explore the design space in detail.
Although the patchy spheres of our model clearly lack the complexity of the units
involved in monodisperse self-assembly in biological systems,
the comprehensive understanding that this simplicity allows us
to achieve should provide useful insights into biological examples.
Furthermore, this feature may be an advantage when considering the design
of synthetic self-assembling building blocks.

The target structure used in this work was relatively simple, and formation of monodisperse clusters
was correspondingly rapid. In future work we intend to explore how the ease of assembly
depends on the geometry of the target structure, and to investigate the limits on the complexity of structures
that can be generated from this relatively simple interaction scheme. In addition, there are various features which
could be added to the potential, most notably a dependence on torsional angles. We expect that a torsional dependence would
greatly increase the range of achievable target structures, and also produce behaviour more closely mirroring that of biological systems.

\begin{acknowledgments}
The authors are grateful for financial support from the EPSRC, the Royal Society
and the Ram\'{o}n Areces Foundation.
\end{acknowledgments}

\end{document}